\documentclass{article}
\usepackage[T1]{fontenc} % add special characters (e.g., umlaute)
\usepackage[utf8]{inputenc} % set utf-8 as default input encoding
\usepackage{ismir,amsmath,cite,url}
\usepackage{graphicx}
\usepackage{booktabs}
\usepackage{tabularx}
\usepackage{color}

\def\lbd{}	% Flag to use correct LBD settings in the paper, please do not modify this line

\title{M\MakeLowercase{u}SFA: Improving Music Structural Function Analysis with Partially Labeled Data}

\oneauthor
{Ju-Chiang Wang \hspace{0.5cm} Jordan B.L. Smith \hspace{0.5cm} Yun-Ning Hung}
{SAMI, ByteDance \\
{\tt\small \{ju-chiang.wang, jordan.smith, yunning.hung\}@bytedance.com}}

\def\authorname{J.-C. Wang, J. B. L. Smith, and Y.-N. Hung}

\begin{document}

\maketitle
\begin{abstract}
Music structure analysis (MSA) systems aim to segment a song recording into non-overlapping sections with useful labels. Previous MSA systems typically predict abstract labels in a post-processing step and require the full context of the song. By contrast, we recently proposed a supervised framework, called  ``Music Structural Function Analysis'' (MuSFA), that models and predicts meaningful labels like `verse' and `chorus' directly from audio, without requiring the full context of a song. However, the performance of this system depends on the amount and quality of training data. In this paper, we propose to repurpose a public dataset, HookTheory Lead Sheet Dataset (HLSD), to improve the performance. HLSD contains over 18K excerpts of music sections originally collected for studying automatic melody harmonization. We treat each excerpt as a partially labeled song and provide a label mapping, so that HLSD can be used together with other public datasets, such as SALAMI, RWC, and Isophonics. In cross-dataset evaluations, we find that including HLSD in training can improve state-of-the-art boundary detection and section labeling scores by \textasciitilde 3\% and \textasciitilde 1\% respectively.
\end{abstract}
\section{Introduction}
\label{sec:introduction}

\begin{figure}[t] \centering
\includegraphics[width=\columnwidth]{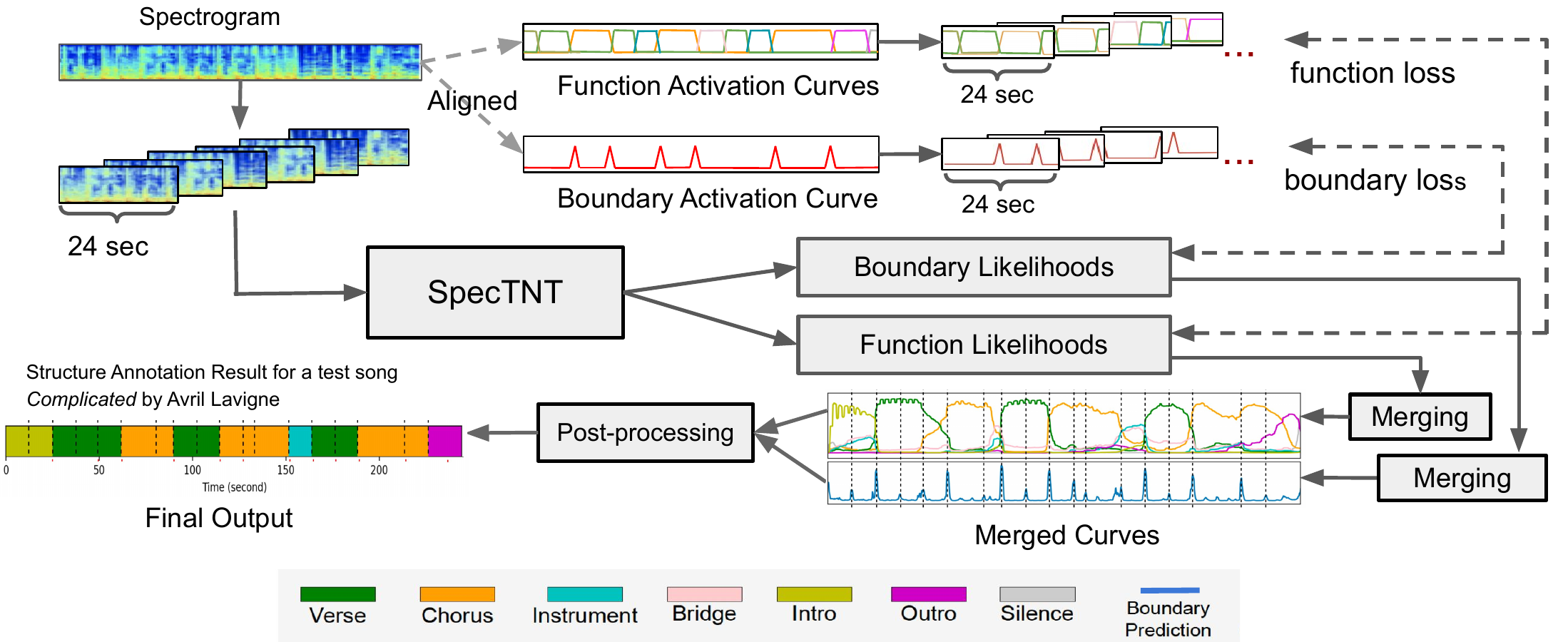}
\caption{Diagram of the DeepMuSFA system.} \label{fig:diagram}
\end{figure}

Music structure analysis (MSA) is often posed as an effort to understand the self-similarity patterns within a piece of music~\cite{nieto2020audio}:
the aim is to identify spans in a piece that repeat and points that represent boundaries between dissimilar regions.
The self-similarity patterns then guide the segment labelling: abstract labels (`A', `B', or `C') or meaningful labels (`chorus', `bridge', etc.) are typically predicted in a post-processing step.
By contrast, in~\cite{wang2022catch} we introduced a supervised ``Music Structural Function Analysis'' (MuSFA) system that predicts meaningful labels directly from audio: using a neural network, it estimated the ``chorusness'', ``bridgeness'' and several other likelihoods as functions of time. The pipeline of the method is illustrated in Figure~\ref{fig:diagram}; please see~\cite{wang2022catch} for details.

This approach has a novel benefit: the system is able to predict the structure of musical excerpts without using the rest of the song as context. E.g., given an excerpt containing the last 3 bars of a verse and the first 5 bars of a chorus, the system could predict the boundary location and segment labels.
A method that relies on modelling the relationship of this excerpt to the rest of the song---to determine which spans repeat, how closely, and how often---would not work in this example.

However, the approach has a cost: it requires a large training dataset. Annotating music structural functions is a laborious task, and the datasets are scarce: four large public ones---Harmonix (912), RWC-Pop (100), Isophonics (277) and SALAMI-pop (274)---total just 1,563 full songs. However, if the system does not require full songs, but can be trained on excerpts, then other data sources could be used. In this paper, we use the HookTheory Lead Sheet Dataset (HLSD)~\cite{yeh2021automatic} to improve an MSA system for the first time. 

The music theory website HookTheory \cite{anderson_hooktheory} invites users to annotate the lead sheet (melody, chords, key and mode) of songs.
However, lead sheets on the site are not for full songs, but for musically meaningful sections like `chorus', `verse', or `intro'.
HLSD has been used in several MIR tasks, including symbolic music generation and harmonization~\cite{yeh2021automatic,jiang2020transformer,chen2021surprisenet}, audio key estimation~\cite{jiang2019mirex}, and melody and chord recognition of audio~\cite{donahue2021sheet}. To our knowledge, this paper is the first work to leverage the structural information of HLSD to help MSA.

\section{Data Pre-processing for HLSD}\label{sec:method}

For this research work, we use an open-sourced collection,\footnote{\scriptsize\url{https://github.com/wayne391/lead-sheet-dataset}} which contains 18,843 music sections, each has the starting and ending times of the section in the corresponding recording.
Two steps are needed to pre-process this data:
(1) map the structural labels into 7 classes, so they can be merged with other datasets; (2) include random lengths of context when excerpting the sample from the full song.

\begin{table}[t]
\centering
\centering
\begin{tabularx}{\columnwidth}{X|l}
\toprule
 \textbf{HLSD labels} & \textbf{MuSFA}  \\
 \hline
chorus, chorus-lead-out, theme, verse-and-chorus, theme-recap, pre-chorus-and-chorus
& chorus  \\
\hline
verse, development, verse-and-pre-chorus, pre-chorus
& verse \\
\hline
instrumental, lead-in-alt, lead-in, loop, solo
& inst \\
\hline
bridge, variation & bridge \\
\hline
intro, intro-and-chorus, intro-and-verse & intro \\
\hline
outro, pre-outro & outro \\
\bottomrule
\end{tabularx}
\caption{HLSD-to-MuSFA label mapping.}
\label{table:mapping}
\vspace{.2cm}	
\end{table}

After parsing the HLSD metadata, we derive the taxonomy that includes about 22 different labels. Table \ref{table:mapping} lists the mapping from a HLSD label (left column) into 6 of the 7 structural labels (right column) used in MuSFA~\cite{wang2022catch}.

Next, to create useful training examples for the model to learn the boundaries,
we select excerpts with a random amount of context, choosing between 8 and 12 seconds of context to include before and after the section.
For instance, suppose that a song has a chorus section that occurs from 0:35–0:58 in the song. Then, we randomly select 9 and 10 seconds of front and rear padding, respectively, meaning we choose the span from 0:26–1:08.
This would result in a 42-second excerpt, within which the span 0:09–0:32 has the label `chorus’ and the remainder was unlabelled.
If the selected context padding exceeds the boundaries of the full song, it is cut: i.e., if the song above ended at 1:00, then the rear padding would be cut to 2 seconds.
We also set a minimum duration of 30 seconds: if the section and the random padding results in an excerpt shorter than that, the padding is extended in either direction until the requirement is met.

\section{Evaluation}\label{sec:experiments}

We treat each excerpt as an independent song with partially annotated boundaries and structural labels.
There are two types of target activation curve defined for our model \cite{wang2022catch}: \emph{boundary activation} and \emph{function activation} curves. For boundary activation, it follows the previous work \cite{wang2021supervised}. For function activation, however, 
we need to alter the \emph{function loss} to only count errors within the boundaries, since we do not know the function labels for the front and rear padding spans of the excerpt.

\subsection{Configuration}
Our goal is to compare the performance of the MuSFA system introduced in~\cite{wang2022catch} using and without using HLSD.
We use four public datasets following the train-test configurations described in \cite{wang2022catch}:
\emph{Harmonix Set} \cite{nieto2019harmonix}, \emph{SALAMI-pop} \cite{smith2011design}, \emph{RWC-Pop} \cite{goto2002rwc}, and \emph{Isophonics} \cite{mauch2009omras2}. 
Four evaluation metrics are used and calculated using the \texttt{\footnotesize mir\_eval} package \cite{raffel2014mir_eval}: (1) \emph{HR.5F}: F-measure of hit rate at 0.5 seconds; (2) Accuracy (\emph{ACC}): the frame-wise accuracy between the predicted function label and the converted ground-truth label; (3) \emph{CHR.5F}: F-measure of `chorus' boundary hit rate at 0.5 seconds; (4) \emph{CF1}: F-measure of pair-wise frames for `chorus' and `non-chorus' sections \cite{wang2021supervised}.

\begin{table}[t]
\centering
%\tiny
\resizebox{\columnwidth}{!}{
\begin{tabular}{l|cccc} 
\toprule
 & HR.5F & ACC & CHR.5F & CF1  \\
 \midrule
 \multicolumn{5}{c}{\textbf{Four-Fold Cross-Validation}} \\
 \midrule
  \multicolumn{5}{c}{~~~~\textit{Harmonix Set}} \\
[0.1cm]
DSF + Scluster~\cite{wang2021deepstruc} & .497 & - & .326 & .611 \\
CNN-Chorus~\cite{wang2021supervised} & - & - & .371 & .692 \\
MuSFA-24s & .570 & .701 & .501 & .815 \\
MuSFA-24s (HLSD) & \textbf{.595} & .714 & \textbf{.512} & .820 \\
MuSFA-36s & .558 & .723 & .476 & .831 \\
MuSFA-36s (HLSD) & .582 & \textbf{.731} & .495 & \textbf{.835} \\

 \midrule
 \multicolumn{5}{c}{\textbf{Cross-Dataset Evaluation}} \\
 \midrule
 
  \multicolumn{5}{c}{\textit{SALAMI-pop}}  \\
[0.1cm]
%DSF + Scluster~\cite{wang2021deepstruc}  & .447 & - & .272 & .573 \\
MuSFA-24s & .490 & .544 & .357 & .811 \\
MuSFA-24s (HLSD) & \textbf{.532} & \textbf{.551} & \textbf{.399} & \textbf{.820} \\
[0.2cm]
  \multicolumn{5}{c}{\textit{RWC-Pop}}  \\
[0.1cm]
%DSF + Scluster~\cite{wang2021deepstruc} & .438 & - & .343 & .653\\
MuSFA-24s & .623 & 675 & .465 & .847 \\
MuSFA-24s (HLSD) & \textbf{.643} & \textbf{.677} & \textbf{.496} & \textbf{.850} \\
[0.2cm]
  \multicolumn{5}{c}{\textit{Isophonics}} \\
[0.1cm]
MuSFA-24s & .590 & .550 & .401 & .733 \\
MuSFA-24s (HLSD) & \textbf{.598} & \textbf{.559} & \textbf{.411} & .\textbf{741} \\

\bottomrule
\end{tabular}}
\caption{Experimental results.}
\label{table:result}
\end{table}

\subsection{Result and Discussion}

Table \ref{table:result} shows the result and comparison. Our baseline systems are named ``MuSFA-24s'' and ``MuSFA-36s'' which use the SpecTNT model \cite{spectnt} with audio chunks of 24 seconds and 36 seconds, respectively. We conduct two types of evaluations: first, we use \emph{Harmonix Set} in a 4-fold cross-validation manner, but with \emph{SALAMI-pop}, \emph{RWC-Pop}, and \emph{Isophonics} included in the training set of every fold.
Second, in cross-dataset evaluations, each of \emph{SALAMI-pop}, \emph{RWC-Pop}, and \emph{Isophonics} in turn serves as the test set, and the remaining datasets (including \textit{Harmonix Set}) are used for training.
For all of these evaluations, we train the MuSFA system as described, or with the addition of HLSD.

We found that adding HLSD can improve performance in all metrics, on average increasing the boundary detection performance (see HR.5F and CHR.5F) by 3\% and the section labeling performance by 1\% (see ACC and CF1). This result is expected: the HLSD example provides clearer context for boundaries; whereas, since the labels of neighboring sections are unknown, the improvement for section labeling is relatively minor.

% For bibtex users:
\bibliography{structure}

\end{document}